\documentclass[10pt, a4paper]{article}

\usepackage{graphicx}
\usepackage{amsmath}
\usepackage{amsfonts}
\usepackage{amsthm}
\usepackage{mathtools}
\usepackage[font=small,labelfont=bf]{caption}
\usepackage{braket}
\usepackage{nicematrix}
\usepackage{subcaption}
\usepackage{siunitx}
\usepackage{cite}
\usepackage[hidelinks]{hyperref}
\usepackage{xcolor}
\hypersetup{
    colorlinks,
    linkcolor={red!80!black},
    citecolor={blue!80!black}
}
\usepackage{abstract}
\usepackage{booktabs}

\usepackage[margin=2.2cm]{geometry}

\title{\Large Ambiguity Clustering: an accurate and efficient decoder for qLDPC codes}
\author{Stasiu Wolanski\thanks{\texttt{stasiu.wolanski@riverlane.com}} \and Ben Barber\thanks{\texttt{ben.barber@riverlane.com}}}
\date{\normalsize Riverlane, Cambridge, UK
\\[0.4cm]
\large 3 January 2025}

\newcommand{\defn}{\emph}

\newcommand{\subfiglab}[1]{\textit{\textbf{#1}}}

\newtheorem{theorem}{Theorem}[section]

\theoremstyle{definition}
\newtheorem{definition}[theorem]{Definition}

\begin{document}

\newcommand{\pluseq}{\mathrel{+}=}

\newcommand{\codeSpace} {\mathcal{C}}
\newcommand{\hilbertSpace}{\mathcal{H}}
\newcommand{\stabSet} S
\newcommand{\singleStab} s
\newcommand{\pauliGroup} {\mathcal{P}}
\newcommand{\singlePauli} g
\newcommand{\numErrors} n
\newcommand{\numChecks} m
\newcommand{\numLogicals} k
\newcommand{\distance} d
\newcommand{\PCM} H
\newcommand{\singPCM} h
\newcommand{\LM} L
\newcommand{\PM} G
\newcommand{\depSet} D
\newcommand{\depVec}{\mathbf{d}}
\newcommand{\rowWeight} w
\newcommand{\colWeight} v
\newcommand{\errOp} E
\newcommand{\eVec} {\bm{e}}
\newcommand{\eVecComp} e
\newcommand{\synVec} {\bm{s}}
\newcommand{\logVec} {\bm{\lambda}}

\newcommand{\tannerGraph}{\mathcal{G}}
\newcommand{\checks}{\mathcal{C}}
\newcommand{\ch}{c}
\newcommand{\errors}{\mathcal{E}}
\newcommand{\err}{e}
\newcommand{\problem}{g}
\newcommand{\logOffset}{\lambda_0}
\newcommand{\logicals}{\mathcal{L}}
\newcommand{\lgl}{\mathfrak{l}}
\newcommand{\checkEdges}{\mathcal{A}}
\newcommand{\logicalEdges}{\mathcal{B}}
\newcommand{\cluster}{\mathcal{K}}

\newcommand{\post} p
\newcommand{\prior} \epsilon
\newcommand{\osdOrder} t
\newcommand{\ACEE} \kappa
\newcommand{\extraEdges}{K}

\newcommand{\errConSet} {\mathcal{E}_{\text{cons.}}}
\newcommand{\chConSet} {\mathcal{C}_{\text{cons.}}}

\newcommand{\syndrome}{\sigma}
\newcommand{\logical}{\lambda}

\newcommand{\rowspan}[1]{\operatorname{row\ span}(#1)}
\twocolumn[
    \maketitle
    \centering
    \begin{minipage}{0.85\textwidth}
    \small
Error correction allows a quantum computer to preserve states long beyond the decoherence time of its physical qubits.
Key to any scheme of error correction is the decoding algorithm, which estimates the error state of qubits from the results of syndrome measurements.
The leading proposal for quantum error correction, the surface code, has fast and accurate decoders,
but several recently proposed quantum low-density parity check (qLDPC) codes allow more logical information to be encoded in significantly fewer physical qubits.
The state-of-the-art decoder for general qLDPC codes, \mbox{BP-OSD}, has a cheap Belief Propagation stage, followed by linear algebra and search stages which can each be slow in practice.
We introduce the Ambiguity Clustering decoder (AC) which, after the Belief Propagation stage, divides the measurement data into clusters that can be decoded independently.
We benchmark AC on the recently proposed bivariate bicycle qLDPC codes and find that, with 0.3\% circuit-level depolarising noise, AC is up to $27\times$ faster than BP-OSD with matched accuracy.
Our implementation of AC decodes the 144-qubit Gross code in \qty{135}{\micro\second} per round of syndrome extraction on an M2 CPU, already fast enough to keep up with neutral atom and trapped ion systems.
    \end{minipage}
\vspace{0.6cm}
]
\saythanks
    
\noindent
Quantum computing has the potential to enable breakthrough calculations in high impact fields including pharmaceutical~\cite{drugsReview} and material~\cite{materialsReview} simulation and cryptography~\cite{shorsAlgorithm}. 
Practical advantage in these areas, however, will only come with fault-tolerant quantum computing, where the constituent qubits of the computer are encoded with sufficient redundancy to endure the rapid and inevitable errors to which they are subject. 
Such a fault-tolerant system requires an encoding scheme---a \defn{quantum code}---for the qubits, in which the state of some number of \defn{logical qubits} is encoded in a larger number of physical qubits. 
The scheme also prescribes sets of measurements to be performed on the physical qubits, from which one can estimate whether an error has occurred and, if it has, determine an appropriate correction,
a calculation referred to as \defn{decoding}.
For practical reasons, it is preferred that each set of measurements involves only a small number of qubits, and each qubit is involved in only a small number of measurements.
A \defn{quantum low-density parity check} (qLDPC) code family is one for which these quantities are bounded by a constant as the number of physical qubits grows large.

The  most widely studied quantum code is Kitaev's surface code~\cite{kitaev}. 
The surface code is qLDPC, with each measurement involving at most four qubits, and each qubit involved in at most four measurements.
In addition, the pattern of measurement allows surface codes to be implemented on a flat surface using a square grid of qubits and nearest-neighbour interactions.
This is a useful property for systems such as superconducting circuits where qubits are implemented as immovable circuit elements. 
The surface code also has extremely fast decoders.
It was remarked in Kitaev's original paper that decoding can be reduced to the problem of finding minimum-weight perfect matchings in a graph.
This can be solved exactly in polynomial time by Edmonds' Blossom algorithm~\cite{blossom}, and more recently approximated in almost-linear time by the Union-Find decoder~\cite{unionFind}.
In practice, the Sparse Blossom implementation PyMatching~2~\cite{pymatching2} can decode a distance 17 surface code experiencing 0.1\% circuit-level noise in less than \qty{1}{\micro\second} per round of measurements, fast enough to keep up with a superconducting quantum processor~\cite{google-below-threshold}.

Bravyi et al.~\cite{IBMCodes} recently proposed implementing a code in the two-block group algebra (2BGA) family~\cite{Kovalev_2013,lin2023quantumtwoblockgroupalgebra} which can protect 12 logical qubits against $0.1\%$ circuit-level noise for one million rounds of syndrome extraction using 288 physical qubits.
The surface code by contrast requires 241 physical qubits to provide only a single logical qubit with a slightly lower level of protection, a greater than $10\times$ increase in the number of physical qubits per logical qubit.
Avoiding the high qubit overhead of the surface code has motivated extensive research into more general qLDPC codes~\cite{QLDPCReview, sqrtLDPC, expanderCodes, asymptoticallyGoodLDPC,jeronimo2021explicitabelianliftsquantum,breuckmann2021balanced,panteleev2021quantum,wang2024coprimebivariatebicyclecodes,xu2023constantoverheadfaulttolerantquantumcomputation,shaw2024loweringconnectivityrequirementsbivariate,berthusen20242dlocalimplementationquantum,neutralAtoms,Ostrev2024classicalproduct,leverrier2022quantum,mostad2024generalizingquantumtannercodes}.

There remain several obstacles to the widespread adoption of non-surface qLDPC codes.
Proposals for computing on the encoded data~\cite{cross2024improvedqldpcsurgerylogical,cohen2022low} are less well-developed than the lattice surgery protocols for the surface code~\cite{latticeSurgery}.
The circuits implementing the codes will generally require more than nearest neighbour interactions on a square grid~\cite{tradeoffsLDPC, Baspin2022connectivity}.
We also lose access to efficient matching-based decoders.

Specialised decoders exist for certain families of codes~\cite{leverrier2022efficient,gu2023efficient,gu2024single,dinur2022goodquantumldpccodes,panteleev2024maximallyextendablesheafcodes}, but the state of the art decoder for general qLDPC codes is Belief Propagation followed by Ordered Statistics Decoding (BP-OSD)~\cite{BPOSDOriginal}.
Belief Propagation (BP) is a classical inference algorithm which effectively decodes classical LDPC codes such as those used in the 5G network~\cite{BP5G}. 
OSD is a post-processing step which handles the fact that quantum codes generally have multiple good explanations for any given set of measurement results (\defn{quantum degeneracy}). 
BP-OSD suffers from a high computational complexity, as it involves both an expensive linear algebra step, cubic in the size of the system, and a search step which is in the worst case exponential in the size of the system.
Measurement data must be processed by the decoder as fast as it is generated, so a decoder which can't keep up requires~\cite{backlog} elaborate `windowing' schemes to process data non-sequentially and in parallel~\cite{parallel-window, alibaba}, complicating overall system design.
Additionally, slow decoders limit our ability to investigate new qLDPC codes.

In this paper we propose a new decoding algorithm, Ambiguity Clustering (AC), for decoding general qLDPC codes.
AC works by forming clusters in a graph describing the code, guided by the (potentially ambiguous) output of BP. 
These clusters can be explored independently, so that the cost of the search step scales only with the size of the clusters, not with the overall size of the decoding problem.

\section*{Ambiguity Clustering}

\begin{figure*}
    \centering
    \includegraphics[scale = 1]{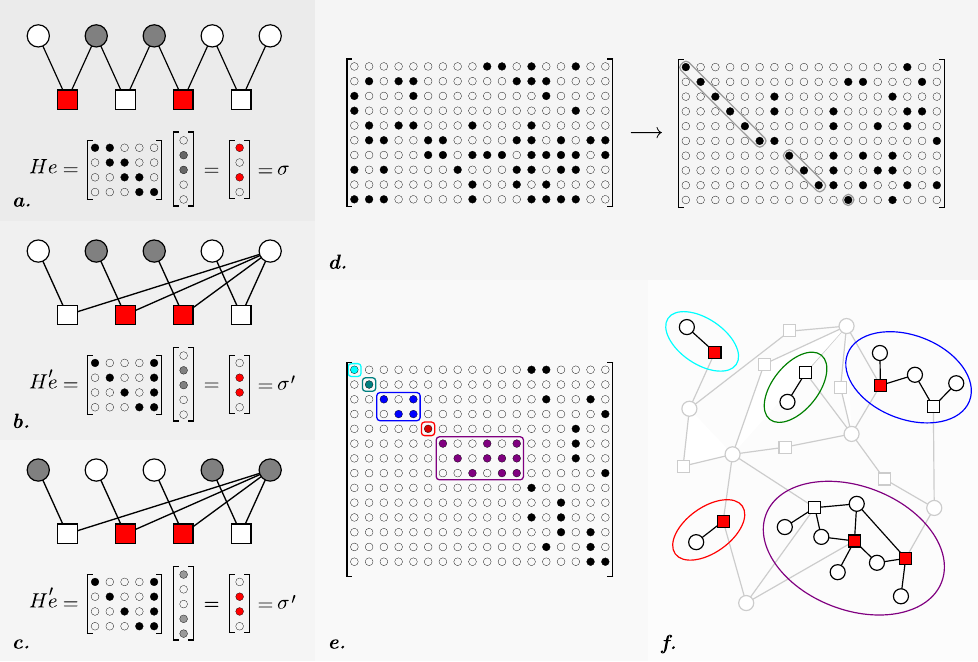}
    { \phantomsubcaption \label{fig:5-bit-rep} }
    { \phantomsubcaption \label{fig:5-bit-rep-alt-1} }
    { \phantomsubcaption \label{fig:5-bit-rep-alt-2} }
    { \phantomsubcaption \label{fig:reduced-form} }
    { \phantomsubcaption \label{fig:block-structure} }
    { \phantomsubcaption \label{fig:cluster-structure} }
    \caption[]{ 
    \textit{The decoding problem and main innovation of AC.}
    \subfiglab{a.} A Tanner graph for the 5-bit repetition code. 
Check nodes (squares) are each connected to an adjacent pair of error nodes (circles), representing the condition that adjacent bits of the codewords $00000$ and $11111$ agree.
An unknown error pattern (shaded circles) has an observable syndrome (shaded squares) consisting of the checks adjacent to an odd number of errors.
Below is the associated parity check matrix $H$.
The syndrome $\sigma$ is the mod 2 sum of the columns corresponding to the unknown error.
Any $e$ satisfying $He = \sigma$ is an explanation of the observed syndrome.

    {}\quad\subfiglab{ b, c.}\ A different choice of Tanner graph and parity check matrix.
The checks represent the conditions that each bit of a codeword should agree with the final bit.
It is particularly easy to read off the sets of columns summing to the observed syndrome: columns 2 and 3 (\subfiglab{b}), or column 5 together with columns 1 and 4 (\subfiglab{c}), corresponding to $e=(0,1,1,0,0)$ and $e=(1,0,0,1,1)$.

{}\quad\subfiglab{d.}\ 
An arbitrary parity check matrix (left) can be converted to a similar special form (right) by Gaussian elimination.
There are typically a great many combinations of columns to search through.
BP-OSD streamlines this search by reordering the columns according to the output of BP, placing columns more likely to represent an error further to the left, before performing Gaussian elimination.
High accuracy can then be obtained by searching over only small sets of columns from the right-hand side of the matrix.

{}\quad\subfiglab{e.}\ 
A partial block structure, covering the rows where the syndrome is non-zero, which would be useful for decoding.
If we are confident that errors corresponding to the rightmost block of columns did not occur,
then we can search over explanations for each section of the syndrome separately.
The space explored by this search can be exponentially larger, in the number of blocks, than a search which does the same amount of work without regard to the block structure.

{}\quad\subfiglab{f.}\ 
Blocks correspond to isolated clusters in an updated Tanner graph.
The key idea of AC is that, by careful manipulation of the parity check matrix, this block structure can usually be obtained.
}
\label{fig:intro-overview}
\end{figure*}

Information about the errors inside a quantum computer is available only indirectly, in the form of a \defn{syndrome} which indicates which of some set of \defn{parity checks} are violated.
Each syndrome can be explained by multiple different underlying errors.
The decoding problem is that of mapping observed syndromes to the most likely underlying errors.

Figure~\ref{fig:intro-overview} gives an overview of the decoding problem and the main innovation of AC.
A decoding problem can be specified by either a \defn{Tanner graph} or a \defn{parity check matrix}.
A Tanner graph has two types of nodes, check nodes and error nodes, with an edge between a check node and an error node whenever the corresponding check is sensitive to the corresponding error (Figure~\ref{fig:5-bit-rep}).
The parity check matrix $\PCM$ is the adjacency matrix of the Tanner graph: $\PCM_{ij}$ is~$1$ if the $i$th check is sensitive to the $j$th error, and $0$ otherwise.
Given a syndrome $\syndrome$, we seek the most likely underlying error vector $e$ such that
\begin{equation}\label{eqn:error-to-syndrome}
\PCM e = \syndrome \pmod 2.
\end{equation}

\defn{Belief Propagation} (BP) is an efficient algorithm to produce estimates $p_j$ for the updated \defn{posterior probability} that $\err_j = 1$ (the $j$th error occurred), given \eqref{eqn:error-to-syndrome}.
BP works by passing messages over the associated Tanner graph.

Classical decoding problems are non-degenerate, meaning
there is typically only one likely solution to \eqref{eqn:error-to-syndrome}.
Provided this solution is sufficiently likely, we can find it by rounding off the posterior probabilities:
\begin{equation}
    e_j = 
    \begin{cases}
    0 & \text{if }p_j < 1/2, \\
    1 & \text{if }p_j \geq 1/2.
\end{cases}
\end{equation}

\begin{figure}
    \centering
    \includegraphics[scale = 1]{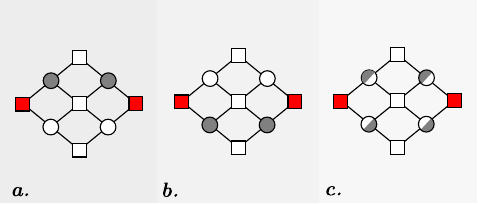}
    \caption[]{\textit{The split belief problem.}
    The syndrome marked in red has two explanations \subfiglab{a}, \subfiglab{b}.
    If all four errors (circular nodes) have the same prior probability, then
    their posterior probabilities are identical (in fact, equal to $1/2$) (\subfiglab{c}).
    It is impossible to retrieve a solution to~\eqref{eqn:error-to-syndrome} by applying a uniform threshold to the posteriors.
    }
    \label{fig:split-belief}
\end{figure}

This is not the case for quantum decoding problems.
There may be many likely solutions to \eqref{eqn:error-to-syndrome} with an identical effect on the stored logical information, so there is no need to distinguish between them.
 The degeneracy of quantum codes nonetheless poses a problem for BP-based decoders, as looking at the most likely value of each $e_j$ individually might not produce an error $e$ compatible with the observed syndrome (see Figure~\ref{fig:split-belief}).
This is known as the \defn{split belief} problem, and configurations leading to this situation are known as \defn{trapping sets}~\cite{splitBelief, trappingSets}.

Several approaches have been proposed for making effective use of BP in quantum error correction. 
These fall into two broad families.
The first is to run BP multiple times, modifying the problem between each round, for example by modifying the prior weights $\prior$, the syndrome $\syndrome$ or the Tanner graph itself, so that a valid global solution can be obtained~\cite{splitBelief, checkAgnosia,stabiliserInactivation,guidedDecimation}.
 
The second, and the one we take here, is to use the BP posteriors to inform a second decoding algorithm which produces the final solution.
This family includes BP-OSD~\cite{BPOSDOriginal} and Belief-Matching~\cite{beliefMatching1, beliefMatching2}.

Gaussian elimination (see Section~\ref{sec:Gaussian-elimination}) is a standard algorithm that converts a parity check matrix to \defn{reduced form} (Figures~\ref{fig:5-bit-rep-alt-1},~\ref{fig:5-bit-rep-alt-2}).
When a parity check matrix is in reduced form  it is easy to read off solutions to \eqref{eqn:error-to-syndrome}, but the number of solutions is typically exponentially large in the problem size.
Most of these solutions will be highly unlikely, requiring many more errors to occur during circuit execution than we expect.
Without further guidance, finding likely solutions $e$ therefore quickly becomes prohibitively expensive as the problem size increases.
BP with Ordered Statistics Decoding post-processing (BP-OSD) uses the output of BP to direct the search to the most promising parts of the solution space.
In BP-OSD, we first sort the columns of the parity check matrix according to the output of BP, so that columns corresponding to the most likely errors are on the left of the matrix.
Then we perform Gaussian elimination, proceeding left to right. 
This converts the matrix to reduced form. 
If the output of BP is reliable, then the best solutions to \eqref{eqn:error-to-syndrome} will largely comprise columns in the resulting identity block, so that accurate results can be obtained by considering only relatively few combinations of the remaining columns.

Broadly speaking, AC works by performing incomplete Gaussian elimination on $\PCM$, with modified pivot selection rules and modified stopping criteria.
This has two advantages: first, it is cheaper than full Gaussian elimination. 
Second, the incomplete Gaussian elimination imposes a block structure on $\PCM$, which allows for a more efficient search over the solution space.
We emphasise that, as the parity check matrix $\PCM$ is transformed, we can reinterpret it as a transformed Tanner graph, and the intuition for many of the steps of AC comes from thinking about this transformed Tanner graph.

After running BP to obtain estimated posteriors~$p_j$, AC proceeds in three stages, which we detail in Section~\ref{sec:AC}.
In the first stage, we transform the Tanner graph to obtain an initial solution to $\PCM \err = \syndrome$. 
At every step of this first stage, we form a simple \defn{cluster} consisting of just one check node and one error node.
By the end of the stage, every \defn{marked} check (corresponding to a $1$ in the syndrome) is in such a cluster.
This allows us to find an initial solution by choosing, for each marked check, the error in its associated cluster.

In the second stage, we grow outwards from this initial solution to identify additional errors, not present in the initial solution, that could be relevant to solving the decoding problem.
During this stage, we grow and merge the simple clusters formed in stage 1 into larger clusters.
This gives us access to a larger space of solutions to look through for each cluster.

In the third stage, we analyse each of the clusters separately to find the likely effect the error in each cluster had on the stored logical information.
We show that for most clusters this logical effect can be calculated very cheaply.
For the remaining \defn{ambiguous clusters}, we perform an approximate maximum-likelihood search to determine the most likely logical effect.
Combining the logical effect of all the clusters gives us an estimate of the overall effect on the stored logical information.

\section*{Performance comparison}

\begin{figure}[t!] 
\centering
  \centering
  \includegraphics[scale = 0.95]{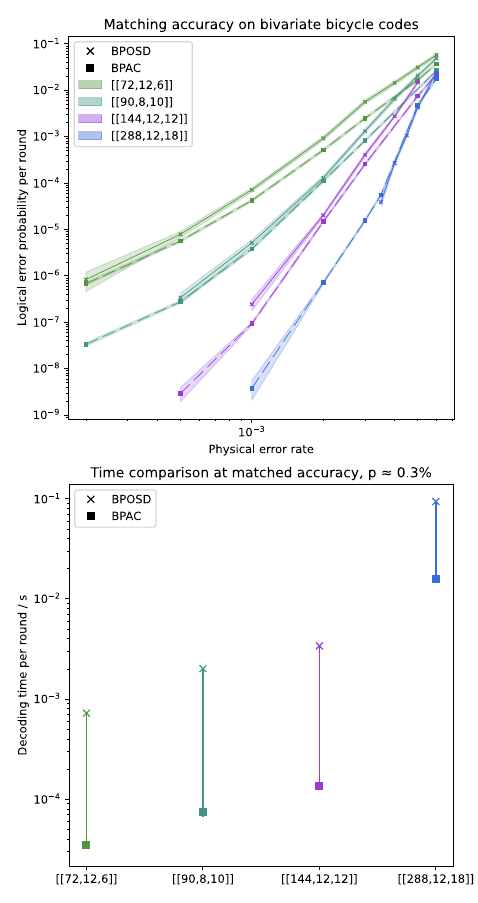}
    \caption[]{
    \textit{Comparison of BP-OSD and AC.}
    
    {}\quad
    \textit{Top}. Logical error rates for quantum memory using the bivariate bicycle codes, labelled by code parameters $[[n,k,d]]$ (the number of physical qubits, logical qubits, and the distance of the code).
    Syndromes were extracted for $\distance$ rounds; logical error and timing data are per round of syndrome extraction.
  Dashed lines are the BP-OSD-CS(7) (see Section~\ref{sec:BPOSD}) data from~\cite{IBMCodes}. 
  For each data point, the AC parameter $\extraEdges$ (Section~\ref{sec:ac-stage-2}) has been set to match or exceed the accuracy of BP-OSD.
  The shaded region indicates one (binomial) standard deviation of error.  
  Only experiments for which we observed at least 5 failures are shown.
  The [[108,8,10]] code is omitted for clarity as it has similar performance to the [[90,8,10]] code.
  
  {}\quad\textit{Bottom}. Decoding time per round of syndrome extraction at $p = 0.3\%$ ($p=0.35\%$ for the [[288,12,18]] code).

  }
    \label{fig:plots}
\end{figure}

We compare BP-OSD and AC on Bravyi et al.'s recently proposed bivariate bicycle codes~\cite{IBMCodes}.
These are 2BGA codes specified by pairs of two-variable polynomials.
In addition to their low qubit overhead these codes were designed with a view to implementation on a superconducting quantum computer, as the required qubit connections could be implemented by making use of both sides of a 2D plane.
Bravyi et al.\ present data for five codes with data qubit counts ranging from 72 to 288.
When we configure AC to match the accuracy of BP-OSD on these codes, simulated subject to a full circuit-level noise model, we observe up to a $27\times$ speedup, using Roffe's standard implementation of BP-OSD~\cite{Joschka, bpOsdSoftware} as a benchmark.

Figure~\ref{fig:plots} shows that AC can be configured to match the logical accuracy data from~\cite{IBMCodes} for the bivariate bicycle codes. 
(See Sections~\ref{sec:numerics} and~\ref{sec:table} for details of the configuration of the decoders in each case.)
We also plot the decoding times for these codes for a physical error rate of $0.3\%$ or $0.35\%$ (whichever BP-OSD accuracy data is available for) with matched logical fidelities.
We see a speedup of between $6\times$ and $27\times$ for the various codes.
In particular, when decoding the 144-qubit Gross code at this physical error rate, we see a speedup of $25\times$, or an absolute decoding time of \qty{135}{\micro\second} per round on a single CPU, with no loss of logical fidelity.
This is already fast enough for real-time decoding on neutral atom~\cite{neutralAtoms} and trapped ion systems~\cite{trappedIonSPAM}.
Neutral atoms have provided the most plausible demonstration to date of all-to-all qubit connectivity \cite{neutralAtomHarvard}, and so are a likely candidate for the first practical implementation of low-overhead qLDPC codes.

Ambiguity Clustering represents a significant step towards real-time decoding of qLDPC codes on near future hardware, especially on slower qubit types including neutral atom systems.
Real-time decoding on solid-state quantum systems, such as superconducting- and spin-based processors, requires a further 2--3 orders of magnitude improvement in decoding time over the CPU decoder presented here.
Bridging this gap will likely require a combination of algorithmic advances and implementation on specialised computing hardware.

\section*{Related work}

After this work was announced, Hillman et al.~\cite{LSD} announced the Localised Statistics Decoder (LSD), which also seeks to avoid using full Gaussian elimination to solve the decoding problem.
AC is driven by global selection of the error most likely to help explain the syndrome at each step, with error clusters condensing from the set of considered errors as the algorithm proceeds.
By contrast, LSD begins with a cluster of errors around each $1$ in the syndrome and is driven by local choices of the most likely adjacent error to pull into each cluster at each step.
Rather than the single representation of the decoding matrix used by AC, LSD maintains independent copies of the parts of the matrix corresponding to each cluster, with rules for reconciling these representations as clusters grow into each other.
We might say that, where AC is top-down, LSD is bottom-up, which could affect the available strategies for parallelising each algorithm.
The existing,  CPU, implementations of both AC and LSD are fully serial.

\section*{Acknowledgements}

We would like to thank Mark Turner for numerous helpful discussions about software, and Dan Browne, Earl Campbell, Joan Camps, Ophelia Crawford, Maria  Maragkou and Luigi Martiradonna  for their valuable comments on drafts of this manuscript.

\section{The decoding problem} \label{sec:the-decoding-problem}

Ambiguity Clustering can be applied to any problem expressed in terms of parity checks on probabilistically independent error mechanisms.
\begin{definition}\label{def:parity-check}
A decoding problem comprises
\begin{itemize}
\item an $m \times n$ binary \defn{parity check matrix} $\PCM$;
\item a $k \times n$ binary \defn{logical matrix} $\LM$;
\item a length $n$ vector of \defn{prior error probabilities} $\prior$;
\item a length $m$ binary \defn{syndrome} vector $\syndrome$.
\end{itemize}
\end{definition}
By a \defn{binary} matrix or vector we mean that the entries are in the finite field $\mathbb F_2 = \{0,1\}$, with arithmetic performed modulo 2.

Given a length $n$ binary \defn{error} vector $e$, write
\begin{equation}\label{eqn:prior-prob-of-error}
\prior(e) = \prod_{i\colon e_i = 1} \prior_i \cdot \prod_{j\colon e_j = 0} (1-\prior_j)
\end{equation}
for the prior probability of $e$.

\begin{definition}\label{def:maximum-likelihood}
The \defn{maximum likelihood decoder} computes the logical effect vector $\logical$ maximising 
\begin{equation}\label{eqn:maximum-likelihood}
\sum_{e \in \{0, 1\}^n \colon \PCM e = \syndrome,\, \LM e = \logical} \epsilon(e).
\end{equation}
\end{definition}
That is, over all errors $e$ consistent with the syndrome $\syndrome$, what is the most likely value of $\LM e$, weighted according to $\prior$?

For a decoding problem to be well-posed, there must be at least one $e$ such that $\PCM \err = \syndrome$.
To simplify the presentation, we will make the stronger assumption that the rows of $\PCM$ are linearly independent, which guarantees that we can always find such an $e$.
This assumption is not necessary (and the decoder implementations do not require it) but in any case can always be achieved by removing linearly dependent rows from $\PCM$.

Definition~\ref{def:parity-check} models many different types of decoding problem, such as classical linear binary codes and quantum stabiliser codes with stochastic noise and perfect measurement.
Here we focus instead on quantum memory with a circuit-level noise model.
See Section~\ref{sec:circuit-level-h-l} for a detailed description of the construction of $\PCM$ and $\LM$ in this setting.

For quantum error correction $n$ is typically much larger than $\numChecks + k$, so for any $e \in \mathbb F_2^n$, there are a large number (at least $2^{n-m-k}$) of $e'$ with both $\PCM e' = \PCM e$ and $\LM e' = \LM e$; that is, that have identical syndromes and logical effects.
This \defn{degeneracy} is a key difference between quantum and classical codes.

\section{Gaussian elimination and BP-OSD}\label{sec:BPOSD}

The state of the art for qLDPC decoding is BP with \defn{Ordered Statistics Decoding} post-processing (BP-OSD).
BP-OSD uses the output of BP to inform an application of Gaussian elimination,
 a standard linear algebra procedure that transforms a matrix $\PCM$ into a form convenient for solving equations of the form $\PCM e = \syndrome$.

Suppose that $\PCM = \begin{bmatrix} I & B \end{bmatrix}$, where~$I$ is an $m \times m$ identity matrix and $B$ is an arbitrary $m \times (n-m)$ matrix.
If we write $e = \begin{bmatrix}f \\ g \end{bmatrix}$, where $f$ has length $m$ and $g$ has length $n-m$,
then each of the $2^{n-m}$ solutions of 
\begin{equation}\label{eqn:solution-space}
\syndrome = He = \begin{bmatrix} I & B \end{bmatrix} \begin{bmatrix}f \\ g \end{bmatrix} 
          = f + Bg
\end{equation}
can be obtained by first choosing one of the $2^{n-m}$ values for $g$, then setting $f = \syndrome + Bg$.

\begin{definition}\label{def:reduced-form}
A column of a matrix is in \defn{reduced form} if it contains exactly one~$1$.
An $m \times n$ matrix is in reduced form if there is a set of $m$ columns in reduced form with their~$1$s in distinct rows.
\end{definition}
A matrix in reduced form differs from the form $\PCM = \begin{bmatrix} I & B \end{bmatrix}$ only by row and column permutations, so is equally convenient for solving $\PCM e = \syndrome$.
This will be a repeating theme: a special form of $\PCM$ will be presented with a particularly convenient ordering of the rows and columns, but any reordering will work equally well.

\subsection{Gaussian elimination}\label{sec:Gaussian-elimination}

\begin{figure}
    \centering
    \includegraphics[scale = 1]{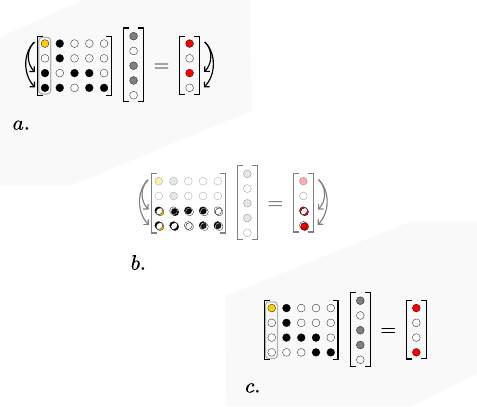}
    \caption[]{\textit{ A pivot operation.} \subfiglab{a.}\ A pivot $ij$ = $11$ (yellow) is chosen.
    \subfiglab{b.}\ The pivot row is added to every other row with a~$1$ in the pivot column. 
    \subfiglab{c.}\ The pivot column is now in reduced form. 
    Further pivot operations in other rows will no longer change this column, so it will remain in reduced form. 
    The syndrome vector on the right is modified in the same way to ensure the solutions (one of which is shown in grey) to the linear system are unchanged.}    
    \label{fig:pivot-operation}
\end{figure}

The basic unit of Gaussian elimination is the \defn{pivot operation} (Figure~\ref{fig:pivot-operation}).
Choose a \defn{pivot row}~$i$ and \defn{pivot column}~$j$ of $\PCM$ that are adjacent (in the corresponding Tanner graph, meaning $\PCM_{ij} = 1$) and add the~$i$th row to every other row $\ell$ adjacent to column~$j$ (with $\PCM_{\ell j} = 1$).
After this pivot operation, the~$j$th column is in reduced form.
Moreover, since the rest of the column is now $0$, further pivot operations will not take the~$j$th column out of reduced form, as long as we avoid pivoting again on the same row.

The description of Gaussian elimination is now very short:
for as long as we can choose a valid pivot entry $ij$ ($H_{ij} = 1$, neither row~$i$ nor column~$j$ has previously been a pivot), pivot at $ij$.
After $m$ pivot operations, $H$ is in reduced form.

We can now decode as follows.
Let $\PCM'$ be a reduced form of $\PCM$ obtained via Gaussian elimination.
Since $\PCM'$ is obtained from $\PCM$ by performing row operations, there is some invertible matrix $R$ recording those row operations such that $\PCM' = R \PCM$.
Given a syndrome $\syndrome$, let $\syndrome ' = R \syndrome$.
Then
\begin{equation}\label{eqn:reduced-form}
\PCM e = \syndrome \iff R\PCM e = R\syndrome \iff \PCM' e = \syndrome'
\end{equation} 
so we have an equivalent decoding problem to the one we started with, but with the parity check matrix in reduced form.
We can now explore the solution space to $\PCM' e = \syndrome'$ using the technique of \eqref{eqn:solution-space}.
From now on, whenever we perform a pivot operation on $\PCM$, we perform the same set of row operations on $\syndrome$ to preserve the solution space to $\PCM e = \syndrome$ (see Figure~\ref{fig:pivot-operation}).
We also omit the primes from $\PCM'$ and $\syndrome'$, using $\PCM$ and $\syndrome$ to refer to the current transformed state of the problem at any point.

\subsection{Ordered Statistics Decoding}

It generally remains impractical to iterate over all $2^{n-m}$ solutions to $\PCM e = \syndrome$.
Searching would be easier if the likely values of $e$ were almost entirely supported on the pivot columns.
\defn{Ordered Statistics Decoding}\label{sec:OSD} (OSD) aims to achieve this by exploiting our freedom to choose a pivot during Gaussian elimination: in each round, we select a pivot $ij$ which maximises $p_j$, BP's estimate  of the probability that $e_j = 1$ given that $\PCM \err = \syndrome$. 
This is typically described as reordering the columns so that $p_j$ decreases left to right, then choosing the leftmost available pivot at each stage.

BP-OSD then is a family of algorithms of the following form.
\begin{enumerate}
    \item Obtain posterior estimates $p_j$ using some form of BP.
    \item Perform Gaussian elimination with the pivot selection rule `maximise $p_j$'.
    \item Search for a solution to the transformed problem $\PCM e = \syndrome$ with highest prior probability $\prior(e)$.
\end{enumerate}

The accuracy and run time can vary depending on the exact implementation of each stage.
We will give a rough indication of the relative cost of each stage.

BP consists of some number of rounds $r$ passing messages along the edges of the Tanner graph. 
The Tanner graph has $m+n$ vertices.
For qLDPC decoding problems each vertex has bounded degree, so there are $O(m+n) = O(n)$
edges, and BP runs in time at most $O(rn)$.
BP is highly amenable to parallelisation, so it is reasonable to think that this could be reduced to $O(r)$ in principle.

Gaussian elimination performs at most $m$ pivot operations, each consisting of at most $m-1$ additions of length $n$ rows, for a total complexity of $O(m^2n)$.
Matrices for qLDPC decoding are sparse, so the practical complexity could be lower even if we make no special effort to use the sparsity, for example if each pivot operation involves relatively few rows.

The cost of searching depends heavily on how thoroughly we search.
There are several named strategies in the literature~\cite{Joschka}.
\begin{itemize}
    \item \defn{Order zero} (BP-OSD-0).
    Take the unique solution which only uses pivot columns.
    That is, in the context of~\eqref{eqn:solution-space}, set $g=0$.
    \item \defn{Exhaustive order $\osdOrder$} (BP-OSD-E($t$)).
    Try all $2^\osdOrder$ values of $g$ supported on the $t$ most likely non-pivot columns.
    \item \defn{Combination sweep order $\osdOrder$} (BP-OSD-CS($\osdOrder$)).
    Try all $g$ of weight at most $1$, and those $g$ with exactly two $1$s in the $\osdOrder$ most likely non-pivot columns.
\end{itemize}

Roffe et al.~\cite{Joschka} report that combination sweep typically gives better accuracy than the exhaustive method searching a similar number of solutions.
BP-OSD-CS($\osdOrder$) considers $1 + (n-m) + \binom \osdOrder 2 = O(n + \osdOrder^2)$ values of $e$.
Scoring each solution by computing $\prior(\err)$ takes time $O(m)$.

For low values of $\osdOrder$, Gaussian elimination is the dominant cost. 
For higher values of $\osdOrder$, we observe that the cost of the solution search dominates. 

Often $\osdOrder$ is taken to be a small constant, giving satisfactory performance on small codes.
However, the existence of trapping sets and the split belief problem gives a heuristic argument that the amount of searching required by BP-OSD should grow exponentially in the size of the problem.
If there are $s$ problematic areas of the syndrome, each of which can be resolved in at least two different ways, then an unstructured search must examine at least $2^s$ solutions.
But if these configurations occur at some constant rate, then for large problems $s$ will scale with $n$ for some constant fraction of syndromes.

The motivation behind AC is to try and identify and resolve these ambiguities independently, turning a multiplicative cost into an additive one.

\section{Ambiguity Clustering} \label{sec:AC}

Here we detail the three stages of AC: stage 1 (Section~\ref{sec:ac-stage-1}), where we find an initial solution; stage 2 (Section~\ref{sec:ac-stage-2}), where we form complex clusters; and stage 3 (Section~\ref{sec:ac-stage-3}), where we analyse these clusters independently.

\begin{figure*}
    \centering
    \includegraphics[scale = 1]{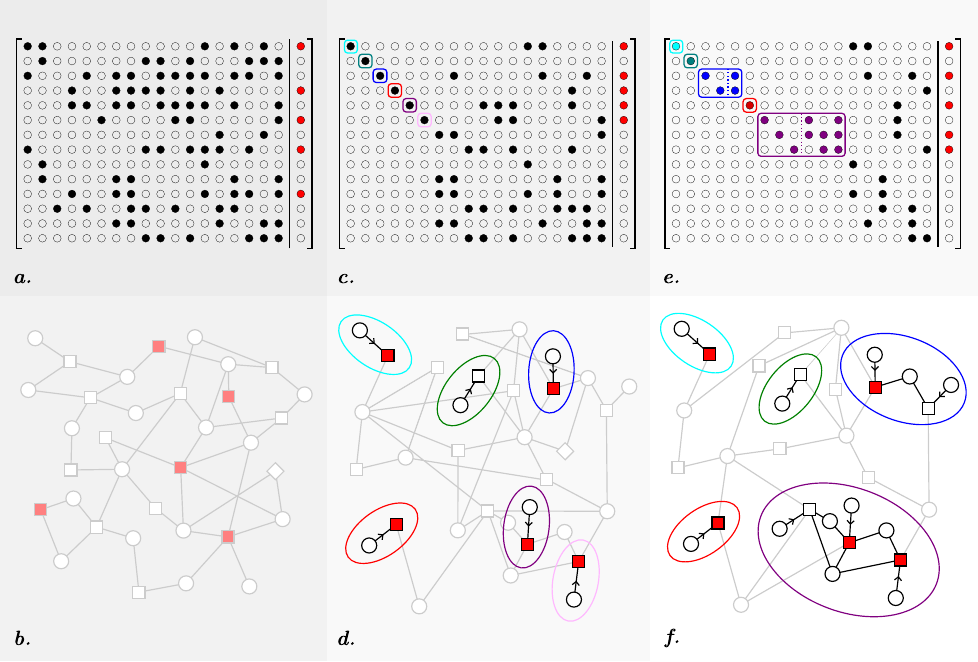}
    { \phantomsubcaption \label{fig:stage-0-H} }
    { \phantomsubcaption \label{fig:stage-0-TG} }
    { \phantomsubcaption \label{fig:stage-1-H} }
    { \phantomsubcaption \label{fig:stage-1-TG} }
    { \phantomsubcaption \label{fig:stage-2-H} }
    {\phantomsubcaption \label{fig:stage-2-TG} }
    \caption[]{
\textit{Evolution of the decoding problem through AC stages~1 and~2.}
    
    {}\quad  \subfiglab{a.} A general, unstructured, parity check matrix $H$.
    The syndrome $\syndrome$ is shown as a column vector on its right.
    
 {}\quad\subfiglab{b.}\ The associated Tanner graph. 
    Checks with a~$1$ in the syndrome are marked in red.
    
  {}\quad  \subfiglab{c.}\ After stage 1, $H$ is reduced with respect to the syndrome (Definition~\ref{def:pcm-solved-form}): an identity block covers the rows in which the syndrome has a $1$. 
  In general, the rows and columns of the identity `block' may be scattered throughout the matrix.
  Because a pivot row can still be involved in later pivot operations, there might (as in this example) be rows in the identity block without a $1$ in the syndrome.
    
  {}\quad  \subfiglab{d.}\ In the Tanner graph, being reduced with respect to $\syndrome$ means that every marked check has an associated error node adjacent only to that check.
    
  {}\quad  \subfiglab{e.}\ After stage 2, an expanded set of blocks covers the syndrome. 
    The blocks $C_i$ are shown as coloured rectangles. 
    Each consists of an $m_i \times m_i$ identity matrix $I_i$, and optionally some linearly dependent columns in the $m_i \times (n_i - m_i)$ matrix $B_i$.
    Again, in practice these rows and columns may be scattered throughout the matrix.
    
  {}\quad  \subfiglab{f.}\ In the Tanner graph, each cluster consists of at least one check node--error node pair, indicated with an arrow, and optionally some additional errors adjacent only to those checks `gluing them together' into a cluster. 
    Every marked check is part of some cluster.
    }
    \label{fig:stages}
\end{figure*}

We provide two figures to help the reader orient themselves.
In Figure~\ref{fig:stages}, we illustrate the evolution of the problem through stages~1 and ~2.
Figure~\ref{fig:flowchart} contains a high-level overview of the algorithm.

\subsection{AC Stage 1: initial solution}\label{sec:ac-stage-1}

Suppose that
\begin{equation}\label{eqn:stage-1-form}
        \PCM = \begin{bmatrix}
        I & \star \\
        0 & \star
    \end{bmatrix}
    \text{ and }
    \syndrome = \begin{bmatrix} \syndrome_1 \\ 0 \end{bmatrix}
\end{equation}
where~$I$ is an $\ell \times \ell$ identity block, $\syndrome_1$ has length $\ell$, a $0$ indicates a block of zeroes, and the blocks indicated by $\star$ have arbitrary entries.
Then one solution to $\PCM e = \syndrome$ is given by
\begin{equation} \label{eqn:ac-zero-sln}
    e
    =
    \begin{bmatrix} \syndrome_1 \\ 0 \end{bmatrix}.
\end{equation}
The goal of stage~1 of AC is to bring the decoding problem into this form.

\begin{definition}\label{def:pcm-solved-form}
Having performed some sequence of pivot operations on distinct rows,
the parity check matrix $\PCM$ is \defn{reduced with respect to $\syndrome$} if every row~$i$ of the matrix with $\syndrome_i = 1$ is a pivot row.
\end{definition}
If $\PCM$ is reduced with respect to $\syndrome$, then we have \eqref{eqn:stage-1-form} up to permutations of the rows and columns.
In comparison to the fully reduced form of Section~\ref{sec:BPOSD},
we have done less work to find the first solution to $\PCM e = \syndrome$.
In return, we no longer have easy access to further solutions.
Recovering these will be the purpose of stage~2.

To reduce $\PCM$ with respect to $\syndrome$, we choose pivots such that each pivot operation makes a row~$i$ with $\syndrome_i = 1$ into a pivot row (possibly at the cost of creating additional~$1$s in the syndrome). 
We break ties using the $p_j$ as in OSD.

Specifically, we iterate the following procedure.
\begin{enumerate}
    \item Choose a pivot row~$i$ and pivot column~$j$ according to the following criteria:
        \begin{enumerate}
            \item the two must be adjacent ($\PCM_{ij} = 1$);
            \item neither has previously been used as a pivot;
            \item \emph{the row must have a non-zero syndrome} ($\syndrome_i = 1$);
            \item of the pairs $ij$ satisfying (a), (b) and (c), choose a pair with maximal $p_j$.
        \end{enumerate}
    \item \label{step:stage-1-pivot-operation} 
      Perform a pivot operation using pivot row~$i$ and pivot column~$j$, updating both $\PCM$ and $\syndrome$ to preserve the solutions to $\PCM e = \syndrome$.
    \item Repeat steps 1 and 2 until there are no more pairs satisfying (a), (b) and (c), at which point $\PCM$ will be reduced with respect to $\syndrome$.
\end{enumerate}

The key difference from OSD is condition 1(c).
In OSD, the order in which pivots are chosen is fixed in advance.
By contrast, here we select the next pivot dynamically based on both $p_j$ and the current state of the updated syndrome.
Since the original syndrome is sparse, this allows the initial solution to be obtained much more quickly than in BP-OSD.

\subsection{AC stage 2: cluster formation}\label{sec:ac-stage-2}

In stage 2 we transform the matrix $\PCM$ into its final form, a block structure that captures the local structure of the solution space.
The blocks can also be viewed as clusters in the associated Tanner graph.

The form we are aiming for is, up to row and column permutations, 
\begin{equation}
    \label{eqn:stage-2-form}
    \PCM =
    \begin{bmatrix}
        C & \star \\ 
        0 & \star
    \end{bmatrix},
\end{equation}
where
\begin{align}\label{eqn:c_i-blocks}
    C = \begin{bmatrix}
          I & B_1 &        &   & \\
            &     & \ddots &   & \\
            &     &        & I & B_c
    \end{bmatrix}.
\end{align}
An example matrix of this form is shown in Figure~\ref{fig:stage-2-H}.
If we reinterpret $C$ as the adjacency matrix of a Tanner graph, the $c$ blocks correspond to connected components.
Figure~\ref{fig:stage-2-TG} shows these components embedded inside the full Tanner graph represented by $\PCM$.

Writing
\begin{equation}
  C_i = \begin{bmatrix}
        I & B_i
    \end{bmatrix},
\end{equation}
each $C_i$ is an $m_i \times n_i$ matrix in reduced form.
Suppose that
\begin{equation}
\syndrome = 
    \begin{bmatrix}
    \syndrome_1 \\
    \vdots\\
    \syndrome_c\\
    0
    \end{bmatrix},
\end{equation}
with each $\syndrome_i$ of length $m_i$.
If we restrict attention to $e$ of the form
\begin{equation}\label{eqn:special-e}
    e = 
    \begin{bmatrix}
    e_1 \\
    \vdots\\
    e_c\\
    0
    \end{bmatrix},
\end{equation}
with each $e_i$ of length $n_i$, then $\PCM e = \syndrome$ if and only if
\begin{equation}\label{eqn:small-systems}
    C_i e_i = \syndrome_i
\end{equation}
for each~$i$.
The restriction \eqref{eqn:special-e} asserts that columns outside $C$ corresponds to error mechanisms which did not occur.
If this is not reliably satisfied, then AC will lose access to solutions that could have been found by OSD.
If it is satisfied, we have split the decoding problem into many smaller decoding problems that can be solved separately.

The partially reduced form \eqref{eqn:stage-1-form} is a trivial example of this block structure with $C$ consisting of the pivot rows and columns, with $m_i = n_i = 1$ throughout.
In stage 2 we will add an additional $\extraEdges$ columns to $C$, and typically also some rows.
Increasing $\extraEdges$ increases the accuracy of the decoder at the cost of increased run time.

\begin{figure}
    \centering
    \includegraphics[scale = 1]{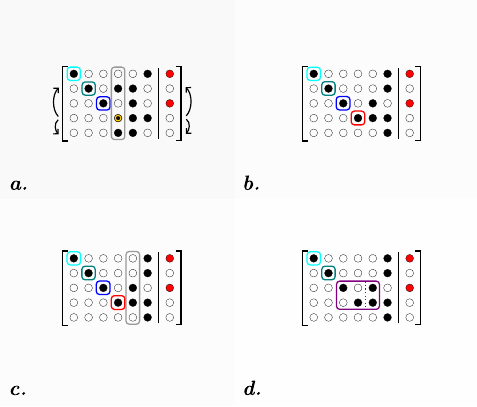}
    \caption[]{\textit{The two stage 2 operations.}
    \subfiglab{a.}\ The selected column has a $1$ in a non-pivot row.
    \subfiglab{b.}\ We perform a pivot operation and seed a new block (red).
    \subfiglab{c.}\ The selected column is supported on pivot rows.
    \subfiglab{d.}\ We add the column to $C$ unchanged, merging blocks (purple) as necessary.
}
    \label{fig:two-options}
\end{figure}

When we add a column $j$ to $C$, one of two things happens, depending on the relationship between that column and $C$ (Figure~\ref{fig:two-options}).

\defn{Seed a new block}.
If column~$j$ has at least one~$1$ outside the rows of $C$, choose such a row~$i$ arbitrarily and pivot at $ij$.
The resulting column~$j$, containing only a single~$1$, can be added as a new block of $C$.
The number of rows, columns and blocks of $C$ each increase by~$1$.

\defn{Grow existing blocks}.
Otherwise, column~$j$ is supported entirely on the rows of $C$, so
we can add it to $C$ unchanged.
It is possible that column~$j$ has~$1$s in multiple blocks of $C$.
In that case, we merge those blocks into a single larger block.
The number of rows is unchanged, the number of columns increases by 1, and the number of blocks either stays the same or decreases.

Stage 2 proceeds as follows.
\begin{enumerate}
    \item Choose a column~$j$ that
    \begin{enumerate}
            \item is not in $C$;
            \item is adjacent to a row that has been involved in a pivot operation in stage 1 or stage 2 (either as a pivot row, or by having a pivot row added to it); and
            \item has maximal $p_j$ among those columns satisfying (a) and (b).
    \end{enumerate}
    \item Add column $j$ to $C$, seeding a new block or growing existing blocks as appropriate.
    \item Repeat steps~1 and~2 until $\extraEdges$ additional columns have been added to $C$.
\end{enumerate}
The goal of step~1 is to select the column most likely to be involved in solutions to $\PCM e = \syndrome$.
This ensures that it is reasonable to restrict attention to $e$ of the form~\eqref{eqn:special-e}, allowing us to solve many small systems~\eqref{eqn:small-systems} separately.

\subsection{AC stage 3: cluster analysis} 
\label{sec:ac-stage-3}

So far, nothing has depended on the logical matrix~$\LM$.
Returning $\LM$ to the picture, 
at the end of stage 2 we have (up to row and column permutations)
\begin{align}
    \PCM & = \begin{bmatrix}
          C_1 &        &     & \star \\
              & \ddots &     & \vdots \\
              &        & C_c & \star \\
            0 & \cdots & 0   & \star     
    \end{bmatrix} & \syndrome = \begin{bmatrix}\syndrome_1 \\ \vdots \\ \syndrome_c \\ 0   \end{bmatrix} \\
    \LM & = \; \begin{bmatrix}
        L_1 & \cdots & L_c & \star 
    \end{bmatrix} \nonumber
\end{align}
where we have labelled by $\LM_i$ the part of $\LM$ restricted to the columns in block $C_i$, and promise to make no further use of the starred columns.

The triples $C_i, \syndrome_i, \LM_i$ express smaller decoding problems which we shall solve separately.
It turns out that many of these problems are easy.

\subsubsection{Unambiguous clusters}

The maximum likelihood solution to the $i$th cluster is the~$\logical_i$ maximising
\begin{equation}\label{eqn:ml-on-cluster}
    \sum_{e_i \in \{0, 1\}^{n_i} \colon C_i e_i = \syndrome_i,\, \LM_i e_i = \logical_i} \epsilon_i(e_i).
\end{equation}
where $\prior_i(e_i)$ denotes the restriction of \eqref{eqn:prior-prob-of-error} to just those columns in $C_i$.

It turns out that there is frequently only one value of $\logical_i$ for which \eqref{eqn:ml-on-cluster} is non-zero.  
That is, there is a unique logical effect consistent with the local syndrome, once we have committed to explaining that part of the syndrome locally.
This is analogous to the neutral clusters arising during execution of the Union-Find decoder~\cite{unionFind}: all corrections within a cluster are logically equivalent, so we can choose between them arbitrarily (for example, by `peeling' in Union-Find).

Unambiguous clusters occur precisely when each row of $\LM_i$ is linearly dependent on the rows of $C_i$.
In that case there is some $k \times m_i$ matrix $R_i$ such that $L_i = R_i C_i$, so that
\begin{equation}
    \lambda_i = \LM_i e_i = R_i C_i e_i = R_i \syndrome_i
\end{equation}
for every $e_i$ with $C_i e_i =  \syndrome_i$;
that is, the logical effect $\logical_i$ is a linear function of the syndrome $\syndrome_i$.

Unambiguous clusters can be detected efficiently: since $C_i$ is in reduced form,
\begin{equation}
    R_i C_i = R_i \begin{bmatrix}I & B_i\end{bmatrix} = \begin{bmatrix}R_i & R_iB_i\end{bmatrix},
\end{equation}
which can only possibly equal $\LM_i$ for a single choice of $R_i$, which is easily checked.

\subsubsection{Ambiguous clusters}

In many cases, all clusters arising from stage 2 are unambiguous and no further analysis is necessary.
An ambiguous cluster must be `large' in the sense that it can be resolved in at least two different ways differing by a logical operation.
We might expect that it would be difficult to arrange two large but disconnected clusters, and when running the decoder we typically see at most one ambiguous cluster.                                                             

The ambiguous clusters are smaller than the original decoding problem, but are still generally too large to allow the maximum likelihood decoding problem to be solved exactly.
We therefore adopt a restricted search strategy, analogous to the search in the final stage of OSD.

As in Section~\ref{sec:Gaussian-elimination}, we let $e_i = \begin{bmatrix} f_i \\ g_i \end{bmatrix}$, with $f_i$ of length $m_i$ and $g_i$ of length $n_i - m_i$, so that the $2^{n_i-m_i}$ solutions to
\begin{equation}
    \syndrome_i = C_ie_i 
                = \begin{bmatrix}I & B_i\end{bmatrix}\begin{bmatrix} f_i \\ g_i \end{bmatrix} 
                = f_i + B_ig_i
\end{equation}
are given by choosing any value for $g_i$ and letting $f_i = \syndrome_i + B_ig_i$.
Then we try the solutions given by
every $g_i$ of weight up to $2$. (Note that BP-OSD-CS tries only a subset of the weight-2 solutions available to it.)
For each solution, we calculate $\prior_i(e_i)$. 
We track the sum of probabilities associated with solutions which flip, and the sum of probabilities of solutions which do not flip, each logical observable $j$. 
If the former sum is greater than the latter, we determine the cluster to have flipped the logical, and we set the $j$th bit of $\logical_i$ to $1$; otherwise, we set it to $0$.

\subsubsection{Combining cluster results}

For each cluster $C_i$, ambiguous or not, we have now calculated an estimate $\logical_i$ for the logical effect of the errors associated with that cluster.
The estimated total logical effect returned by the decoder is $\logical = \sum_{i=1}^c \logical_i$.

Analysing each cluster separately is very efficient.
Considering $s_i$ solutions for each cluster $C_i$ has a cost scaling with $\sum_{i=1}^c s_i$ but covers a part of the solution space to the original decoding problem of size scaling like $\prod_{i=1}^c s_i$.
This is part of what allows us to achieve the accuracy of BP-OSD at much lower computational cost.

\section{Numerical methods}\label{sec:numerics}

Bravyi et al.~\cite{IBMCodes} described circuits to perform syndrome extraction for their codes.
We use versions of these circuits made available by Gong et al.~\cite{guidedDecimation} for Gidney's Clifford simulator Stim~\cite{stim}.
The circuits describe a quantum memory experiment: initialise a block of qubits, extract $r$ rounds of measurement data, then measure the qubits to determine the states of the $k$ logical $Z$ observables.
The decoding problem corresponding to a circuit is described in Section~\ref{sec:circuit-level-h-l}.
As in~\cite{IBMCodes}, we simulate the full syndrome extraction circuit, but only use $Z$-stabiliser measurements to decode the $Z$-logicals.

We use the circuit-level noise model from~\cite{IBMCodes}.
\begin{itemize}
\item \textit{State preparation.}
With probability $p$, prepare the orthogonal state (e.g. $\ket 1$ instead of $\ket 0$).
\item \textit{Measurement.}
With probability $p$, flip the classical measurement result (from $0$ to $1$ or vice versa).
\item \textit{After single qubit gates.}
With probability $p$, apply Pauli $X$, $Y$ or $Z$ with equal probability.
An idle qubit in any time step experiences a noisy identity gate.
\item \textit{After two qubit gates.}
With probability $p$, apply one of the $15$ non-trivial 2-qubit Pauli operations $IX$, $IY$, $IZ$,  $XI$, etc.\ with equal probability.
\end{itemize}

We report the per-round logical failure rate
\begin{equation}
    p_\text{fail} = \frac{N_{\text{fails}}}{rN_{\text{shots}}}
\end{equation}
where $N_{\text{shots}}$ is the number of syndromes decoded, $N_{\text{fails}}$ is the number of times we did not recover the full logical state correctly, and $r$ is the number of rounds of syndrome extraction (equal to the reported distance of the code). 
Second order effects of cancelling logical errors are more complicated to analyse than for codes with a single logical qubit, but also less likely.
We compute $p_\text{fail}$ for BP-OSD in the same way from data kindly provided by the authors of~\cite{IBMCodes}.

Timing data for both AC and BP-OSD were obtained by running each single-threaded on an M2-based MacBook Pro.
A common Python harness generated syndrome data using Stim then called out to either our C++ implementation of AC, or the BP-OSD implementation by Roffe~\cite{Joschka, bpOsdSoftware} in Cython.
Only time spent in the decoder is included in the per-round timing data.

We set the parameters of AC as follows.
For each of 9, 19 and 29 rounds of (sum-product) BP we set $\extraEdges$ (the number of additional columns added to $C$ in stage 2) equal to $\kappa n$, where $n$ is the number of columns of the parity check matrix, and increased $\kappa$ in steps of $0.01$ from $0$ until AC matched or exceeded the accuracy of BP-OSD at each data point plotted in Figure~\ref{fig:plots}.
This gave three timings with comparable accuracy.
The lowest timing was obtained for 9 rounds in almost all cases, so we used 9 rounds of BP throughout.
Where we were able to decode at lower error rates than the authors of~\cite{IBMCodes}, so that there was no BP-OSD comparison data, we maintained the value of $\kappa$ selected for the lowest value of $p$ for which comparison data was available.

The timing data for BP-OSD was obtained using 10 000 rounds of (min-sum) BP and combination sweep order $\osdOrder = 7$, as in~\cite{IBMCodes}.

If BP alone was able to explain the syndrome, then the second stage decoder (OSD or AC) was not used.

In a previous version of this paper we set the number of rounds of BP for both BP-OSD and AC equal to the distance of the code.
This was a bad choice that increased the run times of both BP-OSD and AC.

For BP-OSD, increasing the number of rounds of BP from $d$ to 10 000 decreases the overall running time by around a factor of 5, since it less often has to resort to the expensive OSD step.

Since AC is much faster than OSD, significantly increasing the number of rounds of BP was not advantageous for AC.
Instead, we observed that sum-product BP exhibits strong periodic behaviour, such that the accuracy of AC is significantly increased if we use an odd number of rounds of BP.
We could then trade off this increased accuracy to obtain a (less dramatic) reduction in run time.

\bibliographystyle{unsrt}
\bibliography{bibliography}

\begin{thebibliography}{10}

\bibitem{drugsReview}
Nick~S. Blunt, Joan Camps, Ophelia Crawford, R{\'o}bert Izs{\'a}k, Sebastian
  Leontica, Arjun Mirani, Alexandra~E. Moylett, Sam~A. Scivier, Christoph
  S{\"u}nderhauf, Patrick Schopf, Jacob~M. Taylor, and Nicole Holzmann.
\newblock Perspective on the current state-of-the-art of quantum computing for
  drug discovery applications.
\newblock {\em Journal of Chemical Theory and Computation}, 18(12):7001--7023,
  2022.
\newblock PMID: 36355616.

\bibitem{materialsReview}
Nathalie~P. de~Leon, Kohei~M. Itoh, Dohun Kim, Karan~K. Mehta, Tracy~E.
  Northup, Hanhee Paik, B.~S. Palmer, N.~Samarth, Sorawis Sangtawesin, and
  D.~W. Steuerman.
\newblock Materials challenges and opportunities for quantum computing
  hardware.
\newblock {\em Science}, 372(6539):eabb2823, 2021.

\bibitem{shorsAlgorithm}
P.W. Shor.
\newblock Algorithms for quantum computation: discrete logarithms and
  factoring.
\newblock In {\em Proceedings 35th Annual Symposium on Foundations of Computer
  Science}, pages 124--134, 1994.

\bibitem{kitaev}
A.~Yu {Kitaev}.
\newblock {Quantum computations: algorithms and error correction}.
\newblock {\em Russian Mathematical Surveys}, 52(6):1191--1249, December 1997.

\bibitem{blossom}
Jack Edmonds.
\newblock Paths, trees, and flowers.
\newblock {\em Canadian Journal of Mathematics}, 17:449–467, 1965.

\bibitem{unionFind}
Nicolas Delfosse and Naomi~H. Nickerson.
\newblock Almost-linear time decoding algorithm for topological codes.
\newblock {\em {Quantum}}, 5:595, December 2021.

\bibitem{pymatching2}
Oscar Higgott and Craig Gidney.
\newblock Sparse blossom: correcting a million errors per core second with
  minimum-weight matching, 2023.

\bibitem{google-below-threshold}
Rajeev Acharya, Laleh Aghababaie-Beni, Igor Aleiner, Trond~I. Andersen, Markus
  Ansmann, Frank Arute, Kunal Arya, Abraham Asfaw, Nikita Astrakhantsev, Juan
  Atalaya, Ryan Babbush, Dave Bacon, Brian Ballard, Joseph~C. Bardin, Johannes
  Bausch, Andreas Bengtsson, Alexander Bilmes, Sam Blackwell, Sergio Boixo,
  Gina Bortoli, Alexandre Bourassa, Jenna Bovaird, Leon Brill, Michael
  Broughton, David~A. Browne, Brett Buchea, Bob~B. Buckley, David~A. Buell, Tim
  Burger, Brian Burkett, Nicholas Bushnell, Anthony Cabrera, Juan Campero,
  Hung-Shen Chang, Yu~Chen, Zijun Chen, Ben Chiaro, Desmond Chik, Charina Chou,
  Jahan Claes, Agnetta~Y. Cleland, Josh Cogan, Roberto Collins, Paul Conner,
  William Courtney, Alexander~L. Crook, Ben Curtin, Sayan Das, Alex Davies,
  Laura~De Lorenzo, Dripto~M. Debroy, Sean Demura, Michel Devoret, Agustin~Di
  Paolo, Paul Donohoe, Ilya Drozdov, Andrew Dunsworth, Clint Earle, Thomas
  Edlich, Alec Eickbusch, Aviv~Moshe Elbag, Mahmoud Elzouka, Catherine
  Erickson, Lara Faoro, Edward Farhi, Vinicius~S. Ferreira, Leslie~Flores
  Burgos, Ebrahim Forati, Austin~G. Fowler, Brooks Foxen, Suhas Ganjam, Gonzalo
  Garcia, Robert Gasca, Élie Genois, William Giang, Craig Gidney, Dar Gilboa,
  Raja Gosula, Alejandro~Grajales Dau, Dietrich Graumann, Alex Greene,
  Jonathan~A. Gross, Steve Habegger, John Hall, Michael~C. Hamilton, Monica
  Hansen, Matthew~P. Harrigan, Sean~D. Harrington, Francisco J.~H. Heras,
  Stephen Heslin, Paula Heu, Oscar Higgott, Gordon Hill, Jeremy Hilton, George
  Holland, Sabrina Hong, Hsin-Yuan Huang, Ashley Huff, William~J. Huggins,
  Lev~B. Ioffe, Sergei~V. Isakov, Justin Iveland, Evan Jeffrey, Zhang Jiang,
  Cody Jones, Stephen Jordan, Chaitali Joshi, Pavol Juhas, Dvir Kafri, Hui
  Kang, Amir~H. Karamlou, Kostyantyn Kechedzhi, Julian Kelly, Trupti Khaire,
  Tanuj Khattar, Mostafa Khezri, Seon Kim, Paul~V. Klimov, Andrey~R. Klots,
  Bryce Kobrin, Pushmeet Kohli, Alexander~N. Korotkov, Fedor Kostritsa, Robin
  Kothari, Borislav Kozlovskii, John~Mark Kreikebaum, Vladislav~D. Kurilovich,
  Nathan Lacroix, David Landhuis, Tiano Lange-Dei, Brandon~W. Langley, Pavel
  Laptev, Kim-Ming Lau, Loïck~Le Guevel, Justin Ledford, Kenny Lee, Yuri~D.
  Lensky, Shannon Leon, Brian~J. Lester, Wing~Yan Li, Yin Li, Alexander~T.
  Lill, Wayne Liu, William~P. Livingston, Aditya Locharla, Erik Lucero, Daniel
  Lundahl, Aaron Lunt, Sid Madhuk, Fionn~D. Malone, Ashley Maloney, Salvatore
  Mandrá, Leigh~S. Martin, Steven Martin, Orion Martin, Cameron Maxfield,
  Jarrod~R. McClean, Matt McEwen, Seneca Meeks, Anthony Megrant, Xiao Mi,
  Kevin~C. Miao, Amanda Mieszala, Reza Molavi, Sebastian Molina, Shirin
  Montazeri, Alexis Morvan, Ramis Movassagh, Wojciech Mruczkiewicz, Ofer
  Naaman, Matthew Neeley, Charles Neill, Ani Nersisyan, Hartmut Neven, Michael
  Newman, Jiun~How Ng, Anthony Nguyen, Murray Nguyen, Chia-Hung Ni, Thomas~E.
  O'Brien, William~D. Oliver, Alex Opremcak, Kristoffer Ottosson, Andre
  Petukhov, Alex Pizzuto, John Platt, Rebecca Potter, Orion Pritchard,
  Leonid~P. Pryadko, Chris Quintana, Ganesh Ramachandran, Matthew~J. Reagor,
  David~M. Rhodes, Gabrielle Roberts, Eliott Rosenberg, Emma Rosenfeld, Pedram
  Roushan, Nicholas~C. Rubin, Negar Saei, Daniel Sank, Kannan Sankaragomathi,
  Kevin~J. Satzinger, Henry~F. Schurkus, Christopher Schuster, Andrew~W.
  Senior, Michael~J. Shearn, Aaron Shorter, Noah Shutty, Vladimir Shvarts,
  Shraddha Singh, Volodymyr Sivak, Jindra Skruzny, Spencer Small, Vadim
  Smelyanskiy, W.~Clarke Smith, Rolando~D. Somma, Sofia Springer, George
  Sterling, Doug Strain, Jordan Suchard, Aaron Szasz, Alex Sztein, Douglas
  Thor, Alfredo Torres, M.~Mert Torunbalci, Abeer Vaishnav, Justin Vargas,
  Sergey Vdovichev, Guifre Vidal, Benjamin Villalonga, Catherine~Vollgraff
  Heidweiller, Steven Waltman, Shannon~X. Wang, Brayden Ware, Kate Weber,
  Theodore White, Kristi Wong, Bryan W.~K. Woo, Cheng Xing, Z.~Jamie Yao, Ping
  Yeh, Bicheng Ying, Juhwan Yoo, Noureldin Yosri, Grayson Young, Adam Zalcman,
  Yaxing Zhang, Ningfeng Zhu, and Nicholas Zobrist.
\newblock Quantum error correction below the surface code threshold, 2024.

\bibitem{IBMCodes}
Sergey Bravyi, Andrew~W. Cross, Jay~M. Gambetta, Dmitri Maslov, Patrick Rall,
  and Theodore~J. Yoder.
\newblock High-threshold and low-overhead fault-tolerant quantum memory.
\newblock {\em Nature}, 627(8005):778--782, 2024.

\bibitem{Kovalev_2013}
Alexey~A. Kovalev and Leonid~P. Pryadko.
\newblock Quantum {K}ronecker sum-product low-density parity-check codes with
  finite rate.
\newblock {\em Physical Review A}, 88(1), July 2013.

\bibitem{lin2023quantumtwoblockgroupalgebra}
Hsiang-Ku Lin and Leonid~P. Pryadko.
\newblock Quantum two-block group algebra codes, 2023.

\bibitem{QLDPCReview}
Nikolas~P. Breuckmann and Jens~Niklas Eberhardt.
\newblock Quantum low-density parity-check codes.
\newblock {\em PRX Quantum}, 2:040101, Oct 2021.

\bibitem{sqrtLDPC}
Jean-Pierre Tillich and Gilles Zemor.
\newblock Quantum ldpc codes with positive rate and minimum distance
  proportional to n½.
\newblock In {\em 2009 IEEE International Symposium on Information Theory},
  pages 799--803, 2009.

\bibitem{expanderCodes}
Anthony Leverrier, Jean-Pierre Tillich, and Gilles Zémor.
\newblock Quantum expander codes.
\newblock In {\em 2015 IEEE 56th Annual Symposium on Foundations of Computer
  Science}, pages 810--824, 2015.

\bibitem{asymptoticallyGoodLDPC}
Pavel Panteleev and Gleb Kalachev.
\newblock Asymptotically good quantum and locally testable classical {LDPC}
  codes.
\newblock In {\em Proceedings of the 54th Annual ACM SIGACT Symposium on Theory
  of Computing}, STOC 2022, page 375–388, New York, NY, USA, 2022.
  Association for Computing Machinery.

\bibitem{jeronimo2021explicitabelianliftsquantum}
Fernando~Granha Jeronimo, Tushant Mittal, Ryan O'Donnell, Pedro Paredes, and
  Madhur Tulsiani.
\newblock Explicit abelian lifts and quantum ldpc codes, 2021.

\bibitem{breuckmann2021balanced}
Nikolas~P Breuckmann and Jens~N Eberhardt.
\newblock Balanced product quantum codes.
\newblock {\em IEEE Transactions on Information Theory}, 67(10):6653--6674,
  2021.

\bibitem{panteleev2021quantum}
Pavel Panteleev and Gleb Kalachev.
\newblock Quantum ldpc codes with almost linear minimum distance.
\newblock {\em IEEE Transactions on Information Theory}, 68(1):213--229, 2021.

\bibitem{wang2024coprimebivariatebicyclecodes}
Ming Wang and Frank Mueller.
\newblock Coprime bivariate bicycle codes and their properties, 2024.

\bibitem{xu2023constantoverheadfaulttolerantquantumcomputation}
Qian Xu, J.~Pablo~Bonilla Ataides, Christopher~A. Pattison, Nithin Raveendran,
  Dolev Bluvstein, Jonathan Wurtz, Bane Vasic, Mikhail~D. Lukin, Liang Jiang,
  and Hengyun Zhou.
\newblock Constant-overhead fault-tolerant quantum computation with
  reconfigurable atom arrays, 2023.

\bibitem{shaw2024loweringconnectivityrequirementsbivariate}
Mackenzie~H. Shaw and Barbara~M. Terhal.
\newblock Lowering connectivity requirements for bivariate bicycle codes using
  morphing circuits, 2024.

\bibitem{berthusen20242dlocalimplementationquantum}
Noah Berthusen, Dhruv Devulapalli, Eddie Schoute, Andrew~M. Childs, Michael~J.
  Gullans, Alexey~V. Gorshkov, and Daniel Gottesman.
\newblock Toward a 2{D} local implementation of quantum {LDPC} codes, 2024.

\bibitem{neutralAtoms}
C.~Poole, T.~M. Graham, M.~A. Perlin, M.~Otten, and M.~Saffman.
\newblock Architecture for fast implementation of {qLDPC} codes with optimized
  {R}ydberg gates, 2024.

\bibitem{Ostrev2024classicalproduct}
Dimiter Ostrev, Davide Orsucci, Francisco L{\'{a}}zaro, and Balazs Matuz.
\newblock Classical product code constructions for quantum
  {C}alderbank-{S}hor-{S}teane codes.
\newblock {\em {Quantum}}, 8:1420, July 2024.

\bibitem{leverrier2022quantum}
Anthony Leverrier and Gilles Z{\'e}mor.
\newblock Quantum {T}anner codes.
\newblock In {\em 2022 IEEE 63rd Annual Symposium on Foundations of Computer
  Science (FOCS)}, pages 872--883. IEEE, 2022.

\bibitem{mostad2024generalizingquantumtannercodes}
Olai~Å. Mostad, Eirik Rosnes, and Hsuan-Yin Lin.
\newblock Generalizing quantum {T}anner codes, 2024.

\bibitem{cross2024improvedqldpcsurgerylogical}
Andrew Cross, Zhiyang He, Patrick Rall, and Theodore Yoder.
\newblock Improved {QLDPC} surgery: Logical measurements and bridging codes,
  2024.

\bibitem{cohen2022low}
Lawrence~Z Cohen, Isaac~H Kim, Stephen~D Bartlett, and Benjamin~J Brown.
\newblock Low-overhead fault-tolerant quantum computing using long-range
  connectivity.
\newblock {\em Science Advances}, 8(20):eabn1717, 2022.

\bibitem{latticeSurgery}
Dominic Horsman, Austin~G Fowler, Simon Devitt, and Rodney Van~Meter.
\newblock Surface code quantum computing by lattice surgery.
\newblock {\em New Journal of Physics}, 14(12):123011, 2012.

\bibitem{tradeoffsLDPC}
Sergey Bravyi, David Poulin, and Barbara Terhal.
\newblock Tradeoffs for reliable quantum information storage in 2{D} systems.
\newblock {\em Phys. Rev. Lett.}, 104:050503, Feb 2010.

\bibitem{Baspin2022connectivity}
Nou{\'{e}}dyn Baspin and Anirudh Krishna.
\newblock Connectivity constrains quantum codes.
\newblock {\em {Quantum}}, 6:711, May 2022.

\bibitem{leverrier2022efficient}
Anthony Leverrier and Gilles Z{\'e}mor.
\newblock Efficient decoding up to a constant fraction of the code length for
  asymptotically good quantum codes.
\newblock {\em ACM Transactions on Algorithms}, 2022.

\bibitem{gu2023efficient}
Shouzhen Gu, Christopher~A Pattison, and Eugene Tang.
\newblock An efficient decoder for a linear distance quantum {LDPC} code.
\newblock In {\em Proceedings of the 55th Annual ACM Symposium on Theory of
  Computing}, pages 919--932, 2023.

\bibitem{gu2024single}
Shouzhen Gu, Eugene Tang, Libor Caha, Shin~Ho Choe, Zhiyang He, and Aleksander
  Kubica.
\newblock Single-shot decoding of good quantum {LDPC} codes.
\newblock {\em Communications in Mathematical Physics}, 405(3):85, 2024.

\bibitem{dinur2022goodquantumldpccodes}
Irit Dinur, Min-Hsiu Hsieh, Ting-Chun Lin, and Thomas Vidick.
\newblock Good quantum ldpc codes with linear time decoders, 2022.

\bibitem{panteleev2024maximallyextendablesheafcodes}
Pavel Panteleev and Gleb Kalachev.
\newblock Maximally extendable sheaf codes, 2024.

\bibitem{BPOSDOriginal}
Pavel Panteleev and Gleb Kalachev.
\newblock Degenerate quantum {LDPC} codes with good finite length performance.
\newblock {\em {Quantum}}, 5:585, November 2021.

\bibitem{BP5G}
Yifei Shen, Wenqing Song, Yuqing Ren, Houren Ji, Xiaohu You, and Chuan Zhang.
\newblock Enhanced belief propagation decoder for {5G} polar codes with
  bit-flipping.
\newblock {\em IEEE Transactions on Circuits and Systems II: Express Briefs},
  67(5):901--905, 2020.

\bibitem{backlog}
Barbara~M Terhal.
\newblock Quantum error correction for quantum memories.
\newblock {\em Reviews of Modern Physics}, 87(2):307--346, 2015.

\bibitem{parallel-window}
Luka Skoric, Dan~E Browne, Kenton~M Barnes, Neil~I Gillespie, and Earl~T
  Campbell.
\newblock Parallel window decoding enables scalable fault tolerant quantum
  computation.
\newblock {\em Nature Communications}, 14(1):7040, 2023.

\bibitem{alibaba}
Xinyu Tan, Fang Zhang, Rui Chao, Yaoyun Shi, and Jianxin Chen.
\newblock Scalable surface code decoders with parallelization in time, 2022.

\bibitem{splitBelief}
D.~Poulin and Y.~Chung.
\newblock On the iterative decoding of sparse quantum codes.
\newblock {\em Quantum Information and Computation}, 8:986--1000, 11 2008.

\bibitem{trappingSets}
Nithin Raveendran and Bane Vasi{\'{c}}.
\newblock Trapping sets of quantum {LDPC} codes.
\newblock {\em {Quantum}}, 5:562, October 2021.

\bibitem{checkAgnosia}
Julien du~Crest, Francisco Garcia-Herrero, Mehdi Mhalla, Valentin Savin, and
  Javier Valls.
\newblock Check-agnosia based post-processor for message-passing decoding of
  quantum {LDPC} codes.
\newblock {\em {Quantum}}, 8:1334, May 2024.

\bibitem{stabiliserInactivation}
Julien du~Crest, Mehdi Mhalla, and Valentin Savin.
\newblock Stabilizer inactivation for message-passing decoding of quantum
  {LDPC} codes, 2023.

\bibitem{guidedDecimation}
Anqi Gong, Sebastian Cammerer, and Joseph~M. Renes.
\newblock Toward low-latency iterative decoding of {QLDPC} codes under
  circuit-level noise, 2024.

\bibitem{beliefMatching1}
Ben Criger and Imran Ashraf.
\newblock Multi-path summation for decoding 2{D} topological codes.
\newblock {\em {Quantum}}, 2:102, October 2018.

\bibitem{beliefMatching2}
Oscar Higgott, Thomas~C. Bohdanowicz, Aleksander Kubica, Steven~T. Flammia, and
  Earl~T. Campbell.
\newblock Improved decoding of circuit noise and fragile boundaries of tailored
  surface codes.
\newblock {\em Phys. Rev. X}, 13:031007, Jul 2023.

\bibitem{Joschka}
Joschka Roffe, David~R. White, Simon Burton, and Earl Campbell.
\newblock Decoding across the quantum low-density parity-check code landscape.
\newblock {\em Phys. Rev. Res.}, 2:043423, Dec 2020.

\bibitem{bpOsdSoftware}
Joschka Roffe.
\newblock {LDPC}: {P}ython tools for low density parity check codes, 2022.
\newblock https://pypi.org/project/ldpc/.

\bibitem{trappedIonSPAM}
Fangzhao~Alex An, Anthony Ransford, Andrew Schaffer, Lucas~R. Sletten, John
  Gaebler, James Hostetter, and Grahame Vittorini.
\newblock High fidelity state preparation and measurement of ion hyperfine
  qubits with ${I}>\frac{1}{2}$.
\newblock {\em Phys. Rev. Lett.}, 129:130501, Sep 2022.

\bibitem{neutralAtomHarvard}
Logical quantum processor based on reconfigurable atom arrays.
\newblock {\em Nature}, 626(7997):58--65, 2024.

\bibitem{LSD}
Timo Hillmann, Lucas Berent, Armanda~O. Quintavalle, Jens Eisert, Robert Wille,
  and Joschka Roffe.
\newblock Localized statistics decoding: A parallel decoding algorithm for
  quantum low-density parity-check codes, 2024.

\bibitem{stim}
Craig Gidney.
\newblock Stim: a fast stabilizer circuit simulator.
\newblock {\em {Quantum}}, 5:497, July 2021.

\end{thebibliography}

\appendix

\begin{figure*}
    \centering
    \includegraphics[scale = 0.99]{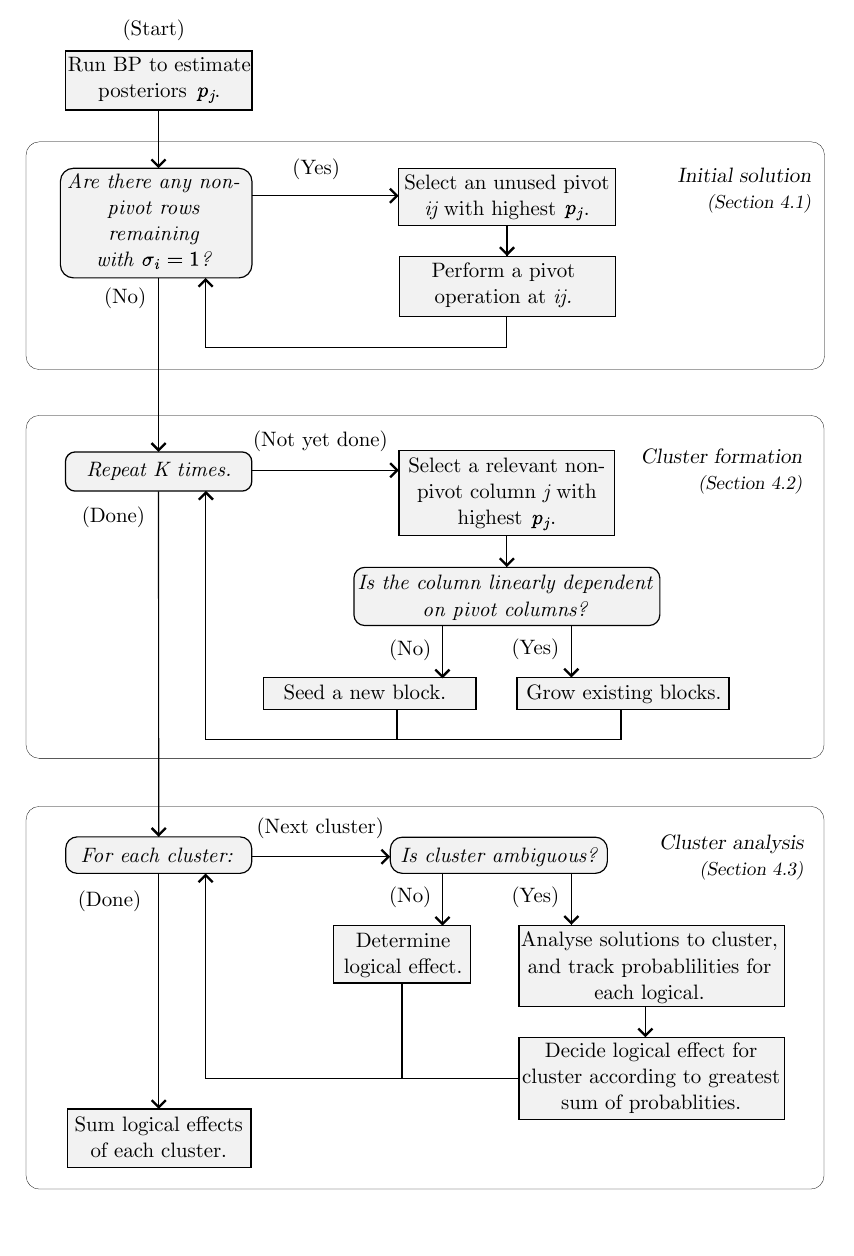}
    \caption{Schematic overview of AC.}
    \label{fig:flowchart}
\end{figure*}

\section{Setting up decoding problems}

In this section we describe how the parity check matrix $H$, logical matrix $L$ and vector of prior probabilities $\prior$ of Section~\ref{sec:the-decoding-problem} can be obtained in various settings.

\subsection{Classical binary linear codes}

Let $H$ be the $(n-k) \times n$ parity check matrix of a classical binary linear code, and let \begin{equation}
G = \begin{bmatrix}
    I & \star
\end{bmatrix}\end{equation} 
be a $k \times n$ generator matrix in standard form.
The logical value of a codeword can be read out by projecting onto the first $k$ coordinates, so the readout of the $i$th encoded bit is affected precisely by errors in the $i$th message bit and we should take
\begin{equation}
    L = \begin{bmatrix}
        I & 0
    \end{bmatrix}
\end{equation}
as the $k \times n$ logical matrix.

Typically the noise model is that each bit flips independently with probability $p$, so $\prior$ will be a constant vector with every entry equal to $p$.

\subsection{Stochastic Pauli noise and perfect stabiliser measurements}

Let $g_1, \ldots, g_m$ be a choice of stabiliser generators for an $n$-qubit stabiliser code and let $q_1, \ldots, q_k$ be a list of logical Pauli operators whose values we wish to decode.
Let $a_1, \ldots, a_{3n} = X_1, Y_1, Z_1, X_2, \ldots, Z_n$ be an enumeration of the non-identity single qubit Pauli errors.
Then the $m \times 3n$ parity check matrix $H$ is defined by
\begin{equation}
    H_{ij} = \begin{cases}
        1 & \text{if } a_j \text{ anticommutes with } g_i, \\
        0 & \text{otherwise.}
    \end{cases}
\end{equation}
Similarly, the $k \times 3n$ logical matrix $L$ is defined by
\begin{equation}
    L_{ij} = \begin{cases}
        1 & \text{if } a_j \text{ anticommutes with } q_i, \\
        0 & \text{otherwise.}
    \end{cases}
\end{equation}

Typically the noise model is thought of as each qubit depolarising independently with some fixed probability $p$. 
As described in Section~\ref{sec:depolarising-as-independent}, this is equivalent to each single qubit Pauli error occurring independently with some other probability $q$, so again $\prior$ is typically some vector with constant entries.

\subsection{Circuit-level noise}\label{sec:circuit-level-h-l}

A \defn{Clifford circuit} is a circuit consisting of qubit preparations in the $X$- or $Z$-bases, Clifford gates, and measurements of Pauli operators.
Clifford circuits are efficiently simulable using the stabiliser formalism, for example using Gidney's Stim~\cite{stim}.

Running (or simulating) a Clifford circuit produces a list of measurement results, which can be identified with a vector in $\mathbb F_2^N$ for some $N$.
Not all measurement sequences in $\mathbb F_2^N$ are achievable: for example, repeated measurement of the same stabiliser will produce the same outcome.
In fact, the achievable measurement sequences form an affine subspace $V$ of $\mathbb F_2^N$.
If we run a Clifford circuit with noise, deviation of the measurement sequence from $V$ is evidence of an error having occurred.

In the context of a quantum memory experiment, the linear equations defining this subspace some in two types.
A \defn{detector} is a set of measurement results whose sum takes a known value under error-free conditions, and which we are allowed to use to diagnose errors.
A \defn{logical observable} is also a set of measurement results whose sum takes a known value under error-free conditions, except that this knowledge relies on us having set up the experiment to preserve a logical $\ket 0$, say, in which case a measurement of the logical $Z$ operator would take a known value.
Logical observables cannot be used for error correction, as in a practical application of quantum error-correction the logical state would be unknown.

For concreteness, we describe the situation for a small quantum memory experiment for a distance 3 rotated surface code with 3 rounds of syndrome extraction.
We might simulate the following Clifford circuit.
\begin{enumerate}
    \item Initialise 9 data qubits in the $\ket 0$ state.
    \item Initialise 4 ancilla qubits in the $\ket 0$ state and 4 in the $\ket +$ state.
    \item Apply the CNOT gates implementing one round of syndrome extraction for the 4 $X$-stabilisers and 4 $Z$-stabilisers.
    \item Measure half of the ancillae in the $X$-basis and half of them in the $Z$-basis.
    \item Repeat steps 2 to 4 twice more.
    \item Measure each data qubit in the $Z$-basis.
\end{enumerate}
This produces a total of $3 \times 8 + 9 = 33$ measurement results.
The 4 first round $Z$-stabiliser measurements are deterministic, so are themselves 4 detectors.
The 4 first round $X$-stabiliser measurement results are random.
However, in the following two rounds the stabiliser measurement results should equal those from the previous round; this gives $8 + 8 = 16$ detectors.
The final round $Z$-measurements are individually random, but certain combinations are deterministic: each final round $Z$-stabiliser measurement is equal to the sum of the $Z$-measurements on its constituent qubits.
This give 4 more detectors, for a total of 24.
There is also one logical observable: the sum of the three final round $Z$-measurements along some appropriate boundary tells us whether we flipped the logical $Z$ operator.

Noise can be added to a Clifford circuit by a variety of error mechanisms.
\begin{itemize}
    \item Preparing the orthogonal basis state, e.g. $\ket 1$ instead of $\ket 0$.
    \item Returning an incorrect measurement result.
    \item Applying a particular Pauli error at a particular location in the circuit.
\end{itemize}
The third type encompasses idle noise and gate errors---see Section~\ref{sec:depolarising-as-independent} for how depolarising a qubit can be recast as the application of independent Pauli errors.
Preparation and measurement errors can also be viewed as certain Pauli error immediately after preparation or before measurement.

Suppose there $m$ detectors, $k$ logical observables, and $n$ error mechanisms in a noisy circuit.
We could set up the decoding problem as follows.
\begin{itemize}
    \item The parity check matrix $H$ is the $m \times n$ matrix whose rows correspond to detectors, columns correspond to error mechanisms, and where $H_{ij} = 1$ precisely if the $i$th detector is flipped by the $j$th error mechanism.
    \item The logical matrix $L$ is the $k \times n$ matrix whose rows correspond to logical observables, columns correspond to error mechanisms, and where $L_{ij} = 1$ precisely if the $i$th logical observable is flipped by the $j$th error mechanism.
    \item The vector $\prior$ records the probabilities with which each of the $n$ error mechanisms occurs.
\end{itemize}
However, it's likely that many of the error mechanisms have an identical effect on both the detectors and the logical observables.  
It is better to combine these error mechanisms to reduce the size of the problem and not force a decoder to choose between them.
For every set of repeated columns, we replace them by a single column with the combined probability that an odd number of them occur.
For example, we replace a pair of repeated columns with probabilities $p_1$ and $p_2$ by a single column with the combined probability $p_1(1-p_2) + p_2(1-p_1)$.
It's also possible that some `errors' have no effect, for example a $Z$ error just before the final $Z$-basis measurement.
This will result in all zero columns, which should be deleted from each matrix.

Another way to reduce the size of the decoding problem, in the case of CSS codes, is to decode $X$ and $Z$ errors separately.
In a quantum memory simulation this means only decoding the error type which can affect the chosen set of logicals.
If we are trying to preserve the $Z$-logicals for example, we can delete the rows of the parity check matrix which only reveal information about $Z$ errors, i.e.\ those corresponding to detectors formed from $X$-stabiliser measurements.
Note that this could produce further identical or zero columns, which will need processing as before.

Taken altogether, this is a reasonably complex process.
Fortunately, Stim can handle most of the complexity for us (including the treatment of zero and repeated columns).
Stim can produce the \defn{Detector Error Model} (DEM) corresponding to a given circuit, which is a list of which sets of detectors and logicals are flipped by each class of error mechanisms, and the probability that an odd number of errors from that class occurs.
The information contained in the DEM is precisely equivalent to that in $H$, $L$ and $\prior$.

\subsection{Phenomenological noise}

The phenomenological noise model is mid-way between circuit-level noise and perfect stabiliser measurement: stabilisers are measured directly, rather than using a short circuit, but still have some probability of being measured incorrectly.
The main advantage of the phenomenological noise model is simplicity: it doesn't rely on finding efficient syndrome extraction circuits, and it is easy to simulate directly.
However, given that circuit-level noise is now so easy to simulate, the easiest way to simulate phenomenological noise is to pick some arbitrary syndrome extraction circuit then use a full Clifford simulator to simulate the circuit with measurement error but without any gate error during the syndrome extraction circuit itself.
The decoding problem can then be specified as in the previous section.

\section{Depolarisation as independent Pauli errors}\label{sec:depolarising-as-independent}

Consider the following two methods for generating a $t$-qubit Pauli error up to phase.
\begin{itemize}
    \item \textit{Depolarising noise of strength $p$.}  
    With probability $p$, select a uniformly random element of the $t$-qubit Pauli group.
    Otherwise, select the identity.
    \item \textit{Independent noise of strength $q$.}  
    Form the product of a random subset of the Pauli group in which each element is included independently with probability $q$. 
\end{itemize}

These appear rather different, as the individual Pauli errors in the first method are mutually exclusive, whereas those in the second method are independent.
In fact, they are completely equivalent.

\begin{theorem}\label{thm:depolarising-as-independent}
For any $0 \leq p \leq 1$, depolarising noise of strength $p$ is identical to independent noise of strength
\begin{equation} \label{eqn:depolarising-as-independent}
    q = \frac{1-(1-p)^\frac{1}{2^{2t-1}}}{2} = \frac{p}{2^t} + O(p^2).
\end{equation}
\end{theorem}

The $t=2$ version of this statement was observed by Gidney (\url{https://algassert.com/post/2001}).

Up to phase, the Pauli group is isomorphic to the vector space $\mathbb F_2^{2t}$ under addition, so for the proof it is convenient to recast the noise processes as follows.
\begin{itemize}
    \item \textit{Depolarising noise of strength $p$.}  
    With probability $p$, select a uniformly random element of $\mathbb F_2^{2t}$.
    Otherwise, select $0$.
    \item \textit{Independent noise of strength $q$.}  
    Form the sum of a random subset of $\mathbb F_2^{2t}$ in which each element is included independently with probability $q$. 
\end{itemize}

\begin{proof}
Let $0 \leq p \leq 1$.
We first claim that there is some $0 \leq q \leq 1/2$ such that the two noise processes agree.

Under depolarising noise of strength $p$, the probability of selecting each non-zero element of $\mathbb F_2^{2t}$ is $p/2^t$.

Under independent noise of strength $q$, the probabilities of selecting each non-zero element of $\mathbb F_2^{2t}$ are all equal by symmetry---any pair of elements are linearly independent, so there is some change of basis which exchanges them.
When $q = 0$, these probabilities are $0$; when $q = 1/2$, they are $1/2^t$
(including each standard basis vector in the sum with probability $1/2$ is already enough to distribute the sum uniformly over $\mathbb F_2^{2t}$).
As $q$ increases from $0$ to $1/2$, this probability varies continuously, so there is some intermediate value of $q$ for which it equals $p/2^t$.

To determine the value of $q$, we compute the probability that the first bit of the random vector is $1$.
For depolarising noise, this probability is $p/2$.

For independent noise, let this probability be $Q$.
Since half of the elements of $\mathbb F_2^{2t}$ have first bit equal to $0$ and half have first bit equal to $1$, $Q$ is the probability that $B_1 + \cdots + B_{2^{2t-1}} = 1$, where $B_1, \ldots, B_{2^{2t-1}}$ are $2^{2t-1}$ independent random samples from $\mathbb F_2$.
Then
\begin{align}
   1-2Q & = (1-Q)(-1)^0 + Q(-1)^1 
\\      & = \mathbb E((-1)^{B_1 + \cdots + B_{2^{2t-1}}})
\\      & = \mathbb E((-1)^{B_1}) \times \cdots \times \mathbb E((-1)^{B_{2^{2t-1}}})
\\      & = (1-2q)^{2^{2t-1}},
\end{align}
where the third equality uses the independence of the $B_i$.
It follows that
\begin{equation}
Q = \frac{1-(1-2q)^{2^{2t-1}}}{2}.
\end{equation}
Equating this to $p/2$ gives
\begin{equation}
1-p = (1-2q)^{2^{2t-1}}
\end{equation}
and so
\begin{equation}
q = \frac{1-(1-p)^\frac{1}{2^{2t-1}}}{2},
\end{equation}
since $0 \leq q \leq 1/2$.
\end{proof}

Note that in this section we used the parameterisation of depolarising noise in which there is a $p/2^t$ probability of applying each of the $2^t-1$ non-trivial Pauli errors.
If the alternative convention is used in which this probability is instead $p/(2^t-1)$ then $p$ must be rescaled appropriately before applying \eqref{eqn:depolarising-as-independent}.

\section{Data in Figure 1}
\label{sec:table}

Table~\ref{fig1data} gives the values, error estimates and decoder parameters for the AC data points plotted in Figure~\ref{fig:plots}.

    The maximum number of rounds of message passing in the Belief Propagation stage was 9 in each case.
    
    The parameter $\kappa$ controls the amount of additional exploration taking place in stage~2---after the completion of stage~1 the decoder explores an additional $\kappa$ fraction of the total number of columns in the parity check matrix.
    
    Logical error per round is the fraction of shots in which any logical qubit was decoded incorrectly, divided by the number of rounds of syndrome extraction.
    In particular, no adjustment is made for the varying number of logical qubits of each code.
    Values are shown as central estimate $\pm$ one standard deviation, assuming that the number of decoding failures was sampled from a binomial distribution with the observed mean.
    
    Decoding time per round is the average number of seconds spent in the decoder, divided by the number of rounds of syndrome extraction.
    The indicated error is the empirical standard deviation of the decoding time divided by the number of rounds of syndrome extraction and the square root of the number of shots.
    
    Standard deviations are given to one significant figure, and means to the same number of decimal places as the standard deviation.
    The final row of the table, for which only one decoding failure was observed, was not plotted in Figure~\ref{fig:plots}.

\begin{table*}
    \centering
    \begin{tabular}{clllcc}
    \toprule
$n$ & $p$  & $\kappa$ & Fails/shots & Logical error per syndrome round & Decoding time per syndrome round \\
\midrule
72 & 0.006    & 0.0      &  251/1130             & $(3.7\pm0.2)\times 10^{-2}     $ & $(2.5\pm0.1)\times 10^{-4}     $\\
 & 0.005    & 0.0      &  251/2550             & $(1.6\pm0.1)\times 10^{-2}     $ & $(1.56\pm0.05)\times 10^{-4}   $\\
 & 0.003    & 0.0      &  251/16891            & $(2.5\pm0.2)\times 10^{-3}     $ & $(8.19\pm0.09)\times 10^{-5}   $\\
 & 0.002    & 0.0      &  251/81702            & $(5.1\pm0.3)\times 10^{-4}     $ & $(4.60\pm0.02)\times 10^{-5}   $\\
 & 0.001    & 0.0      &  251/994627           & $(4.2\pm0.3)\times 10^{-5}     $ & $(3.167\pm0.003)\times 10^{-5} $\\
 & 0.0005   & 0.0      &  251/7441582          & $(5.6\pm0.4)\times 10^{-6}     $ & $(2.6163\pm0.0009)\times 10^{-5}$\\
 & 0.0002   & 0.0      &  251/61766755         & $(6.8\pm0.4)\times 10^{-7}     $ & $(2.2472\pm0.0002)\times 10^{-5}$\\
\midrule
90 & 0.006    & 0.0      &  251/953              & $(2.6\pm0.1)\times 10^{-2}     $ & $(9.1\pm0.5)\times 10^{-4}     $\\
 & 0.003    & 0.0      &  251/30522            & $(8.2\pm0.5)\times 10^{-4}     $ & $(1.158\pm0.009)\times 10^{-4} $\\
 & 0.002    & 0.0      &  251/227814           & $(1.10\pm0.07)\times 10^{-4}   $ & $(8.74\pm0.02)\times 10^{-5}   $\\
 & 0.001    & 0.01     &  251/6614042          & $(3.8\pm0.2)\times 10^{-6}     $ & $(5.623\pm0.002)\times 10^{-5} $\\
 & 0.0005   & 0.01     &  251/90945974         & $(2.8\pm0.2)\times 10^{-7}     $ & $(4.6122\pm0.0004)\times 10^{-5}$\\
 & 0.0002   & 0.01     &  68/408219080        & $(1.7\pm0.2)\times 10^{-8}     $ & $(4.53\pm0.04)\times 10^{-5}   $\\
\midrule
144 & 0.006    & 0.0      &  251/936              & $(2.2\pm0.1)\times 10^{-2}     $ & $(4.8\pm0.3)\times 10^{-3}     $\\
 & 0.005    & 0.0      &  251/2817             & $(7.4\pm0.4)\times 10^{-3}     $ & $(1.60\pm0.07)\times 10^{-3}   $\\
 & 0.003    & 0.0      &  251/81966            & $(2.6\pm0.2)\times 10^{-4}     $ & $(3.10\pm0.02)\times 10^{-4}   $\\
 & 0.002    & 0.01     &  251/1424103          & $(1.47\pm0.09)\times 10^{-5}   $ & $(2.536\pm0.002)\times 10^{-4} $\\
 & 0.001    & 0.03     &  219/195677840        & $(9.3\pm0.6)\times 10^{-8}     $ & $(2.364\pm0.007)\times 10^{-4} $\\
 & 0.0005   & 0.03     &  9/256844300        & $(3\pm1)\times 10^{-9}         $ & $(1.340\pm0.005)\times 10^{-4} $\\
\midrule
288 & 0.006    & 0.0      &  251/783              & $(1.78\pm0.09)\times 10^{-2}   $ & $(9.6\pm0.4)\times 10^{-2}     $\\
 & 0.005    & 0.0      &  251/2938             & $(4.7\pm0.3)\times 10^{-3}     $ & $(2.19\pm0.09)\times 10^{-2}   $\\
 & 0.0035   & 0.05     &  251/251265           & $(5.5\pm0.4)\times 10^{-5}     $ & $(6.14\pm0.01)\times 10^{-3}   $\\
 & 0.003    & 0.05     &  251/1211448          & $(1.15\pm0.07)\times 10^{-5}   $ & $(8.001\pm0.005)\times 10^{-3} $\\
 & 0.002    & 0.05     &  65/11119770         & $(3.2\pm0.4)\times 10^{-7}     $ & $(5.084\pm0.008)\times 10^{-3} $\\
 & 0.001    & 0.05     &  1/17793970         & $(3\pm3)\times 10^{-9}         $ & $(3.021\pm0.005)\times 10^{-3} $\\
 \bottomrule
\end{tabular}
    \caption{The data presented in Figure~\ref{fig:plots}.}
    \label{fig1data}
\end{table*}

\end{document}